\documentclass{webofc}
\usepackage[varg]{txfonts}   % Web of Conferences font
\usepackage{lineno}

\usepackage[figurename=Fig.]{caption}
\begin{document}
\title{New developments for ALICE MasterClasses and the new Particle Therapy MasterClass}

\author{\firstname{Łukasz} \lastname{Graczykowski}\inst{1}\thanks{\email{lukasz.kamil.graczykowski@cern.ch}}, \fnsep\firstname{Piotr} \lastname{Nowakowski}\inst{2}\fnsep\thanks{\email{p.nowakowski@cern.ch}} \and
        \firstname{Panagiota} \lastname{Foka}\inst{3}\fnsep\thanks{\email{Yiota.Foka@cern.ch}}
        % etc.
\\on behalf of the IPPOG Collaboration}

\institute{
 \inst{1}Faculty of Physics, Warsaw University of Technology, Koszykowa 75, 00-662 Warszawa, Poland\\
\inst{2}Faculty of Electronics and Information Technology, Warsaw University of Technology, Nowowiejska 15/19, 00-665 Warszawa, Poland\\
\inst{3}GSI Helmholtzzentrum fur Schwerionenforschung GmbH, Planckstrasse 1, 64291 Darmstatd, Germany
          }

\abstract{%
International MasterClasses (IMC), an outreach activity of the International Particle Physics Outreach Group (IPPOG), has been bringing cutting-edge particle physics research to schoolchildren for over 15 years now. All four LHC experiments participate in the event, including ALICE, the experiment optimised for the study of heavy-ion collisions. Heavy-ion physics is actively contributing to IMC with new developments such as experimental measurements but also applications for society, such as treatment of cancer with ions. In particular, ALICE provides three MC measurements related to the main observables used to characterize the properties of the produced Quark-Gluon Plasma. Historically, those MC measurements were developed independently, inheriting from the first one, by several ALICE groups. Since all of them are based on the ROOT EVE package, a project to integrate them into a common framework was undertaken. ALICE delivers now a single and easy-to-use application, compiled under Linux, MacOS, and, for the first time, Windows. Then, in line with current IPPOG goals to increase the global reach and scope of the IMC programme a newly developed measurement on medical applications of particle physics, the Particle Therapy MasterClass (PTMC) was introduced in the IMC2020 programme. It is a simplified version of matRad, a MATLAB-based toolkit for calculation of dose deposition in the body and allows for planning of radiotherapy using different modalities and highlighting the benefits of treatment with ions.

}
\maketitle
\section{Introduction}
\label{intro}
The International Particle Physics Outreach Group (IPPOG)~\cite{IPPOG} is a collaboration of particle physicists and engineers, science educators and communication specialists from all over the world, with the aim of communicating the fundamental particle physics research to general public. The ``International MasterClasses -- Hands-on Particle Physics'' (IMC)~\cite{IMC} is the leading activity of IPPOG, which aims at providing high school students with access to particle physics data with dedicated packages of analysis software and instructions. Every year in the period from February to April, students are invited to one of the participating universities or research laboratories to attend particle physics lectures and perform measurements using real data from the Large Hadron Collider (LHC)~\cite{Evans:2008zzb} experiments. All four LHC experiments (ATLAS, CMS, LHCb, and ALICE) participate in the IMC providing various measurements on different aspects of particle physics. All MC measurements use especially prefiltered samples of real collision data, recorded by the respective experiment. In most cases, they are built on visualisation tools that are part of the experiment’s software framework. Hence, students have the possibility to work on real data as scientists do. Thousands of students from across the globe attend the event annually. 

Recently, the increase of the global reach and scope of the IMC programme has become one of the main goals of IPPOG. Several new measurements coming from non-LHC and non-CERN experiments are currently in various stages of preparation. Moreover, the expansion of IMC is not limited to fundamental research only. Of particular interest are applications of technologies developed initially for particle physics that our society benefits from. For instance, medicine is the field where such a benefit is undoubtedly the most obviously visible and has a direct impact on our life.  In particular, cancer has recently become the most common cause of death in high-income countries~\cite{DAGENAIS2020785}.

Heavy-ion physics is present within the International MasterClasses since the beginning of the LHC IMC programme, with continuous developments aiming at expanding in scope and reach. In line with the physics research of typical heavy-ion experiments, three ALICE measurements were developed based on actual data analysis of the experiment. They correspond to the most important observables studied to characterise the properties of matter produced in energetic collisions of ions or protons. Then, with the aim to highlight applications from fundamental research for society, the new Particle Therapy MasterClass (PTMC) was introduced in the 2020 edition of IMC for the first time, familiarising students with the actual operation technique used for cancer treatment employing photons, protons or carbon ions. As a good example is presented the fact that for this therapy the heavy-ion research centre GSI, Germany, played a pioneering role in Europe, in the 90s, leading to the construction of the HIT therapy centre in Heidelberg. 

Both MasterClasses had challenging requirements on software developments as they had to be realistic but also easy to use. 

\section{ALICE experiment and ALICE MasterClasses}
\subsection{ALICE}
ALICE (A Large Ion Collider Experiment)~\cite{Aamodt:2008zz} is the LHC~\cite{Evans:2008zzb} experiment dedicated to the study of heavy-ion collisions. The primary objective of ALICE is the study of the properties of the produced deconfined state of matter, the quark-gluon plasma (QGP) which, according to theory, existed in the very beginning of our Universe. A number of observables are studied to fully characterize the hot and dense matter produced in the laboratory in energetic collisions of lead nuclei and systematically compare them with proton-proton collisions of different particle multiplicities. ALICE developed MasterClass measurements for three of the main observables: a) strange particles production, b) nuclear modification factor and c) $\rm J/Psi$ production.

\subsection{ALICE MasterClass}
The ALICE software is based on the ROOT framework~\cite{Antcheva:2009zz} which is used for the experiment’s data processing and analysis. Consequently, the ALICE MasterClasses are also based on ROOT to fulfil the requirement of the MC to be as close as possible to the experimental reality. The advantage is that school-children are thrilled to know that they use the very same tools as the ALICE scientists. The downside is that the installation of the ROOT framework is required which makes the preparation process cumbersome. To bypass this issue Virtual box installation was used. Historically, the MasterClass on strangeness was developed first \cite{ALICE_Masterclass_strangeness} and then adapted for the nuclear modification factor \cite{ALICE_Masterclass_Raa}. Those fully developed and documented MC measurements are included in the IMC schedule. The third MC on the J/Psi measurement was also developed with starting point the existing MC software and is currently being finalised. Despite their common origin they soon diverged as different groups and individuals implemented additional functionality or corrections in the individual computer codes with no tracking of different versions. In order to facilitate maintenance and coherency, a concerted effort started as a summer student project in 2018, followed by an EU-funded grant of the Warsaw University of Technology group, with the aim to unify the existing MC code integrating it in a single framework, taking advantage at the same time of the upcoming enhanced functionality of ROOT 6. The new framework, where all three ALICE MC are integrated, is now used in the IMC 2020 schedule.

\subsection{New developments}

\begin{figure}[ht]
\centering
\sidecaption
\includegraphics[width=0.7\linewidth]{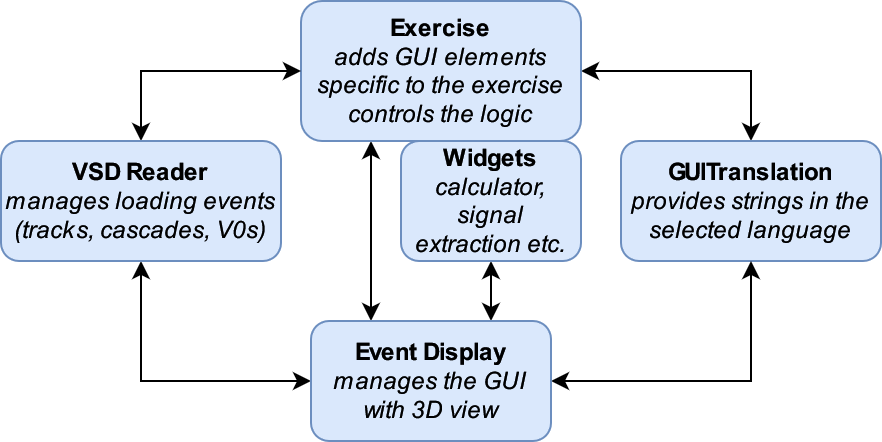}
\caption{Structural diagram of the new ALICE MasterClass framework integrating the three existing measurements.}
\label{diagram}
\end{figure}

Figure~\ref{diagram} represents the new design of the ALICE MasterClass framework at the functional level. The aim was to identify common elements such as the Event Display and reading of experimental data files. Hence, the main component of the program is an instance of \textit{Event Display}, which handles the 3D detector display and track visualisation. Selecting a measurement from the main menu activates one of the \textit{Exercise} objects, which control the scenario of each task. This object can transform the \textit{Event Display}'s default GUI by injecting one or more \textit{Widget} elements, such as particle mass calculator or particle count histogram. In this way the GUI can be dynamically modified to suit a specific measurement. Another common functionality, reading of the experiment data files, is realized by another object --- the \textit{VSD Reader}. The new development also supports translation of the program into other languages, which can be helpful for students not proficient in English. The \textit{GUITranslation} object is responsible for serving the translated text of instructions, error messages, button descriptions \textit{etc.} for both \textit{Exercise} and \textit{Event Display}. 

The new framework is distributed and installed as compiled binary standalone application which presents advantages compared to the previous script-based and interpreted form of the project in terms of size and performance. The new version ROOT 6 was used, replacing ROOT 5. Although ROOT itself is still required for the MasterClass to function, it was possible to bundle it with the application, eliminating the need to install it separately. The ``embedded'' ROOT is not installed in any global directory and will not interfere with any previous ROOT installations on the system. Thanks to the improvements made in this ROOT version, it was possible to port the ALICE MC framework to other major operating systems (both Windows 10 and MacOS X), which hugely improves the accessibility of the program. For the first time, it allows the participating students to use the ALICE MC framework at home on their own computers, regardless of the operating system. In addition, a VirtualBox disk with preinstalled Ubuntu and the ALICE MC is also provided.

\begin{figure}[ht]
\centering
\sidecaption
\includegraphics[width=0.6\linewidth]{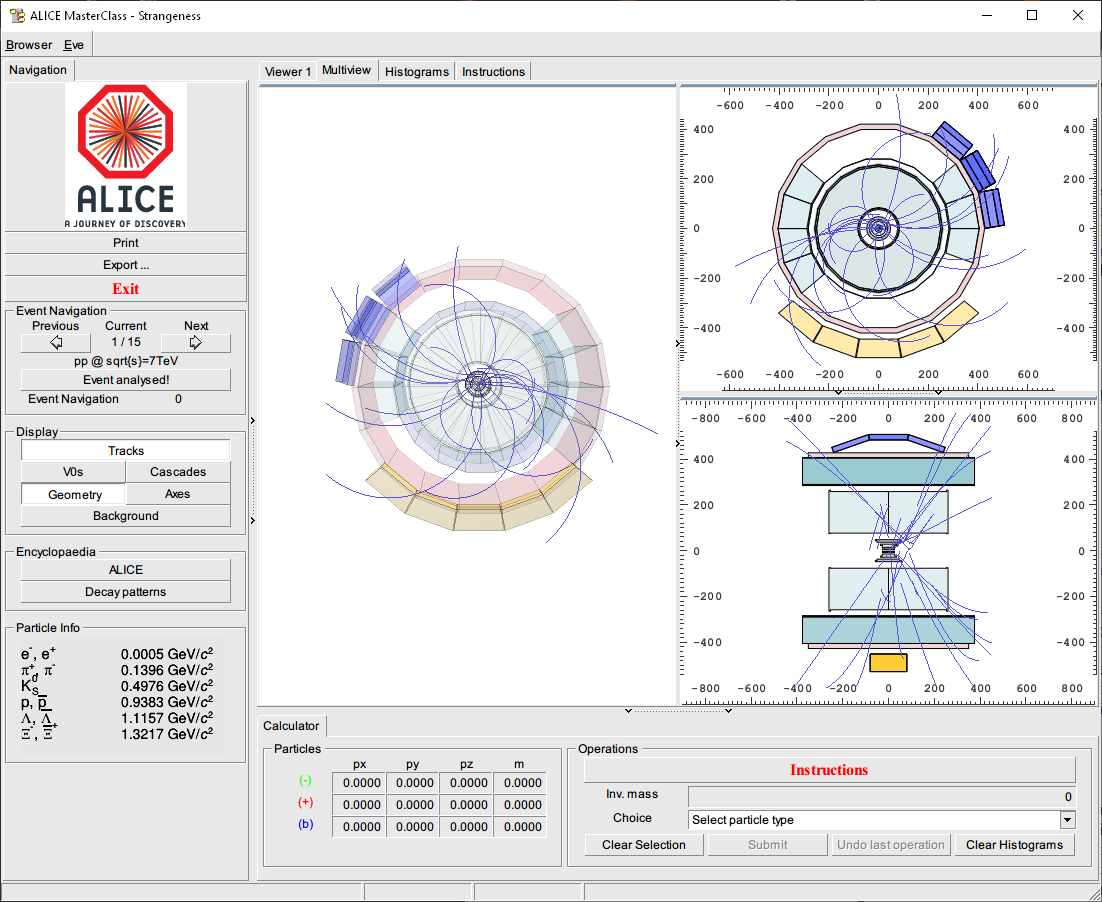}
\caption{Main window of the ALICE MasterClass (production of strange particles measurement) showing the 3D visualisation of the ALICE detector with reconstructed particle tracks from pp collisions at $\sqrt{s}=7$~TeV.}
\label{stg-1}
\end{figure}

To create the package for Linux systems a \textit{AppImage} framework was used, which bundles the whole application (executable, ROOT framework and experimental data) into a single file, which can be executed via a double-click gesture like any other program. For Mac OSX a standard \textit{pkg} installer was created which places the MC among other programs in the \textit{Applications} directory. For Windows a standard \textit{.msi} installer was created which can be used to place the MC app in \textit{Program Files} directory with Desktop and Start Menu shortcuts. Figure~\ref{stg-1} shows the event display of the strangeness MC in the new framework.

The new ALICE MC suite of applications can be downloaded from the official ALICE MC website~\cite{masterclassweb}.

\subsection{Summary of the latest developments}
The list below summarises all recent changes made in the ALICE MasterClass framework:
\begin{itemize}
    \item ALICE MasterClass as an optimized compiled binary, not a set of scripts,
    \item Portable code --- a version available for Windows, Linux and MacOS X,
    \item ROOT bundled with the app, not required to install separately, will not affect other existing installations,
    \item Many bug fixes such as restoring the functionality of system minimize and exit buttons,
    \item Extendable framework for future exercises,
    \item Possible to translate the app into other languages, language selection at runtime.
\end{itemize}

\newpage
\section{Particle Therapy MasterClass}
With the aim to enhance awareness on the benefits for society, and in particular for health, that result from fundamental research, a new MC package was developed: the Particle Therapy MasterClass (PTMC). It is included for the first time in the official schedule of IMC2020 after a very successful pilot in April 2019 with the participation of GSI, CERN and DKFZ Heidelberg~\cite{PTMC_masterclass}. This core team represents the world leading research institutes in their domain that also provided tangible applications for health: at GSI the first 450 patients in Europe were experimentally treated with carbon ions~\cite{HIT_ion_beam}, at CERN the Proton-Ion Medical Machine Study (PIMMS) design team was hosted~\cite{Badano:385378} and DKFZ is the German Cancer research centre next to the HIT Heidelberg ion therapy facility~\cite{DKFZ_masterclass}.

Students have the opportunity to realise that very basic physics principles constitute the basis of treatment of cancer tumours and that very same instruments used in physics research laboratories are also used in cancer therapy facilities. In particular, it is illustrated that as particle accelerators evolve and become more powerful and efficient, their medical applications become more and more important and accessible.

The PTMC is based on the matRAD open-source software~\cite{matrad} which itself is based on MATLAB~\cite{matlab}. It is a professional treatment planning toolkit developed by DKFZ for research and training. It is used to optimize the ``prescription'' of treatment of cancer tumours using photons, protons or carbon ions. Hence, students can appreciate the differences of these modalities and optimally use them according to different cases. While it gives comparable results with commercial treatment planning software actually used for therapies, it is lightweight, flexible and provides results quite fast.

\begin{figure}[ht]
\centering
\sidecaption
\includegraphics[width=1\linewidth]{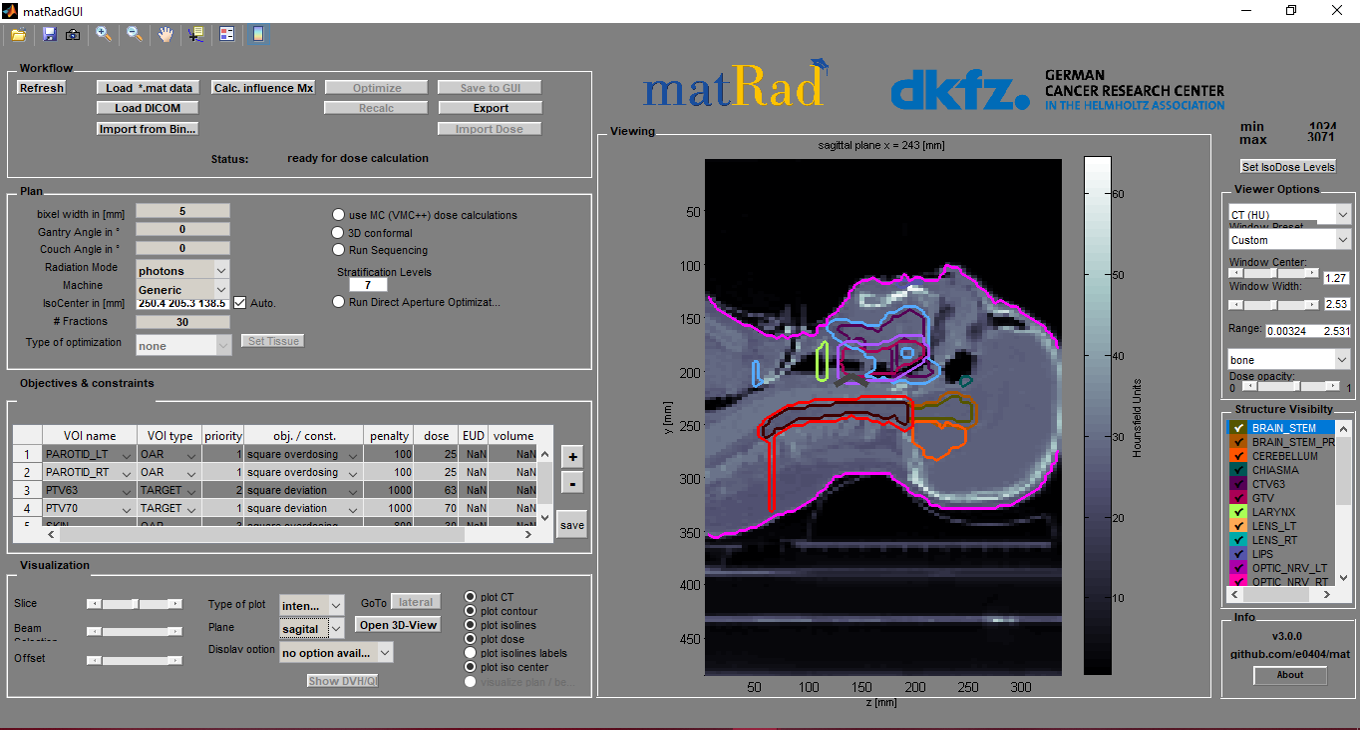}
\caption{Main window of the Particle Therapy Masterclass showing therapy planning for the head tumour case.}
\label{ptmc_gui}
\end{figure}

The data provided are a ``water phantom'' and computer tomography (CT) scans of head and liver tumour cases. matRAD presents the data in 3D, provides possibilities to rotate interactively and present slices in 2D projections. It provides graphical tools to select the volumes to be irradiated allowing for margins, and a particular exercise is planned to demonstrate the importance of accurate alignment. matRAD also provides tools to select the critical organs where the dose should be minimised. Once the volumes to maximise and minimise the dose are selected one can choose the type of radiation (photons, protons or carbon ions) and different angles for irradiation. For the optimisation, one can use constraints and define a minimum and/or maximum dose value. The visual result on the display, based on a colour scale, is very intuitive to comprehend and easy to associate with the resulting histograms of the delivered dose to the target tumour and organs at risk to avoid. Students can see the photon radiation depositing a larger dose in the tissues before the target tumour and penetrating healthy tissues behind the tumour. They can select different angles for irradiation to minimize the deposited dose in each trajectory while accumulating the required dose in the target point. The students can then witness the different properties of protons, and in particular of carbon ions, that, in contrast, deposit very little energy in the tissues before and practically none after the tumour. Figure~\ref{ptmc_gui} shows the PTMC application with the head tumour example.

The accompanying lectures are related to particle interactions with matter, and with radiation in general spanning from its use in microwave ovens to discoveries studying cosmic radiation. The impact of accelerators becomes clear quoting some fifteen thousand used for medical applications, but also the need for further research and optimization. Imaging, diagnostics, dosimetry, are a critical part of the treatment procedures where again the impact of breakthrough developments on detectors for physics experiments is clear. This hands-on experience also makes clear the importance of computing and software developments and usual questions bring the discussion on machine learning techniques and their possible applications for these cases. Students taking part in the pilot session were impressed with the possibility to perform a treatment planning like professionals and motivated to contribute to the further advancements of this multi-disciplinary field. Overall, the obvious need for properly trained scientists was an unavoidable conclusion and take-home message.

All the necessary material is available via the web page~\cite{ptmc_webpage} which includes details for the matRAD installation (on Windows, Linux and MacOS), different workflows together with the associated demos and step-by-step presentations, instructions for tutors and moderators as well as lectures and animations. 

\section{Summary}
The International MasterClasses programme is currently undergoing substantial increase in terms of number of experiments participating and available measurements. Moreover, with the inclusion of the Particle Therapy MasterClass the scope of the programme extends now beyond fundamental research in particle physics.

The ALICE MasterClass framework was redesigned and the three measurements are now integrated into one common framework,  based on ROOT, as the experiment's software. The macro-based solution was replaced by compiled standalone applications for all three major operating systems (Windows, Linux, MacOS X) which are provided with easy to use installers. In addition, the ROOT package was embedded in the framework making transparent its installation. Current works are focused on moving the whole framework into a web-based solution and a client-server architecture, while keeping the possibility to use ROOT-data files and ROOT objects.

The new Particle Therapy MasterClass extending the scope of the International MasterClasses programme to medical applications is prepared following the general rules of other MasterClass measurements, providing a simplified tool based on the actual computer program for treatment planning. Data of liver and head tumour cases and a variety of beam choice (photons, protons, carbon ions) allow for detailed analysis of possible treatment scenarios. The first PTMC session within the IMC2020 was attended by 200 students in Mexico and shows the big interest for measurements on applications.

\section*{Acknowledgements}
The ALICE MasterClass is a part of the ``MatFizChemPW'' project which is supported by the European Union through the European Social Fund under the  Operational Programme Knowledge Education Development (Poland).

\bibliography{bibliography}

\end{document}